\documentclass[aps,pre,showpacs]{revtex4-1}
\usepackage[dvipdfmx]{color,graphicx}
\usepackage{amsmath,amssymb}
\usepackage{algorithm}
\usepackage{algpseudocode}

\begin{document}
\title{Coarse-grained sensitivity for multiscale data assimilation}
\author{Nozomi Sugiura}
\email{nsugiura@jamstec.go.jp}
\affiliation{Japan Agency for Marine-Earth Science and Technology}
\begin{abstract} We show that the
 effective average action and its gradient are useful
 for solving multiscale data assimilation problems.
 We also present a procedure for numerically evaluating
 the gradient of the effective average action,
 and demonstrate that the
 variational problem for slow degrees of
 freedom can be solved properly
 using the ``effective gradient.''
\end{abstract}
\maketitle

\section{Introduction}
A key problem in the  assimilation of data for nonlinear multiscale systems concerns the optimization of the
slow degrees of freedom after the fast degrees of
freedom have been properly averaged~\cite{QJ:QJ837}. This is also the
case with data assimilation for coupled atmosphere--ocean  systems
\cite{npg-11-319-2004}.
From a statistical point of view,
this amounts to performing some integration
with respect to similar realizations of control variables,
and packing them together
into an ``effective'' cost function (action) \cite{Abarbanel20094044}
(see below for the definition).
Geometrically,
the rough surface of the original cost function can be smoothed
according to a coarse-grained averaging procedure.

To see this, we first review
how the concept of
the effective action
is
relevant to
data assimilation \cite{abarbanel2013predicting}.
Data assimilation concerns the following statistical
problem:
given the observation $y \in \mathbb{R}^p$,
the prior probability $P(\chi)$ of the
control variable $\chi \in \mathbb{R}^M$,
 and
 the likelihood $P(y|\chi)$ of the observation,
 the conditional expectation
of any physical quantity $G(\chi)$
 is calculated through the integral:
\begin{align}
E[G(\chi)|y] &= \frac{
\int \mathrm{d}\chi G(\chi) P(\chi) P(y|\chi)
}
{
\int \mathrm{d}\chi P(\chi) P(y|\chi)
}
=
\frac{
\int \mathrm{d}\chi G(\chi) \mathrm{e}^{-S[\chi]}
}
{
\int \mathrm{d}\chi \mathrm{e}^{-S[\chi]}
},\label{expectation}
\end{align}
where
$S[\chi]$
is called the action, or cost function, and
\begin{align}
\int \mathrm{d}\chi=\int_{-\infty}^{\infty} \int_{-\infty}^{\infty} \cdots \int_{-\infty}^{\infty} d\chi_1 d\chi_2 \cdots d\chi_M,
 \end{align}
denotes the multiple integral over
all possible combinations (paths) of $\chi$,
 also called the path integral.
Although the control variable $\chi=\chi(x,t)$ can  generally
be a field defined in some space-time $(x,t)$,
we confine ourselves
to the case of a discrete space-time with $M$ cells,  that is,
$\chi \in \mathbb{R}^M$.
Note that Eq.\,(\ref{expectation}) includes the posterior probability
$P(\chi|y)=P(\chi)P(y|\chi)/P(y)
$
as
a special case with the delta functional
$G(\chi')=\delta(\chi'-\chi)$.

If the posterior $P(\chi|y)$ is highly concentrated
around the most probable state $\hat{\chi}$,
which means $P(\chi|y) \simeq \delta(\chi-\hat{\chi})$,
Eq.\,(\ref{expectation}) can be approximated as:
\begin{align}
E[G(\chi)|y] &\simeq G(\hat{\chi}).
\end{align}
In this situation,
it is important to
find the control variable that
minimizes the cost function $S[\chi]$.
4D-Var efficiently determines one of the stationary points
 satisfying $\delta S[\chi]/\delta \chi = 0$.
However, it does not necessarily give the desired global minimum, 
as the cost function may have multiple minima.
With regard to the shape of the cost function,
we could fail to see the forest for the trees.

To deal with more general posterior probabilities,
and to calculate  the conditional expectation more robustly,
we present an effective alternative to the cost function.
We introduce
an external source term $-J^T \chi$ ($J \in \mathbb{R}^M$ is an external
field) to the action
in the normalization factor
$\int \mathrm{d}\chi \exp{\left(-S[\chi]\right)}$
of Eq.\,(\ref{expectation}).
This leads to the following definition of the partition function:
\begin{align}
Z[J] &= \int \mathrm{d}\chi \mathrm{e}^{-S[\chi] +J^T \chi},\label{def_Z}
\end{align}
which encodes all the information about the conditional expectation as follows:
\begin{align}
E[G(\chi)|y] &= \frac1{Z(0)}\left. G\left(\frac{\delta}{\delta J}\right) Z[J] \right|_{J=0},
\end{align}
where $G(\delta / \delta J)$ should be interpreted as an operator in
which the argument $\chi$ of the
algebraic expression $G(\chi)$ is replaced with the differential operator.
The logarithm of the partition function:
\begin{align}
 W[J] &= \log{Z[J]}
\end{align}
is also useful, because it contains all the information about the
cumulants. For example,
\begin{align}
\frac{\delta W}{\delta J}[0] &=  E[\chi^T|y], \label{cumulant1}\\
\frac{\delta^2 W}{\delta J^2}[0] &=
E\left[\left( \chi- E[\chi|y] \right) \left( \chi- E[\chi|y] \right)^T
 |y \right].\label{cumulant2}
\end{align}
That is to say, we can extract information about the expected value
if we perturb the external field $J$ and observe how the normalization
factor $Z[J]$, or $W[J]$, changes.

To estimate the expected value,
we can construct a functional called the effective action \cite{Abarbanel20094044,abarbanel2013predicting},
whose independent variable is
the expected value $\phi$ in the presence of the external field $J$,
through the Legendre transformation:
\begin{align}
\Gamma[\phi] &\equiv \sup_J{\left\{-W[J] + J^T  \phi \right\} }.
 \label{def_Gamma}
\end{align}
Taking the supremum while $\phi$ remains fixed in Eq.\,(\ref{def_Gamma}), we obtain
 \begin{align}			
   \phi^T = \frac{\delta W}{\delta J}[J].
  \end{align}
Using this and taking the derivative with respective to $\phi$ gives
 \begin{align}			
   \frac{\delta \Gamma}{\delta \phi}[\phi]
&=
-\frac{\delta W}{\delta J}[J]
 \frac{\delta J}{\delta \phi}
+\frac{\delta W}{\delta J}[J]
 \frac{\delta J}{\delta \phi}
+J^T
=J^T.\label{derive_J}
  \end{align}

There are at least two advantages to this transformation.
First, it results in a convex function because we 
take the Legendre transform of a convex function $W[J]$
 \footnote{
The convexity of $W[J]$ follows from 
  H{\"o}lder's inequality:
 $
 E\left[\exp{ \left((1-\gamma) J_1^T\chi+\gamma J_2^T\chi \right)}\right]
 \leq
 E\left[\exp{ \left(J_1^T\chi \right)}\right]^{1-\gamma}
 E\left[\exp{ \left(J_2^T\chi \right)}\right]^{\gamma},~ 0 \leq \gamma \leq 1
 $,
where
  $E\left[ G(\chi) \right] \equiv  
 \int \mathrm{d}\chi G(\chi) \exp{\left(-S[\chi]\right)}
 /
 \int \mathrm{d}\chi  \exp{\left(-S[\chi]\right)}$
}.
Second, it elicits a
symmetric relation between
$W[J]$ and $\Gamma[\phi]$:
  \begin{align}
   \phi^T = \frac{\delta W}{\delta J}[J],
   \quad J^T = \frac{\delta \Gamma}{\delta \phi}[\phi].\label{def_Sym}
  \end{align}
This implies that
its unique stationary point, which satisfies $\delta \Gamma[\phi]/\delta \phi=0$,
identifies the expected value $\phi=\delta W[0]/\delta J=E[\chi|y]$.
In other words, we can find the conditional expectation by
finding the stationary point of the effective action.
Eq.\,(\ref{def_Sym}) also implies
\begin{align}
 \frac{\delta^2 \Gamma}{\delta \phi^2}[\phi]
=
 \frac{\delta J}{\delta \phi}
=
\left(
\frac{\delta \phi}{\delta J}
\right)^{-1}
=
\left(
\frac{\delta^2 W}{\delta J^2}[J]
\right)^{-1}.\label{d2gdphi2}
\end{align}
Comparing this with
(\ref{cumulant2}), we see that the
stationary point of the effective action
also provides more  cumulant information.

Since $\Gamma[\phi]$ in Eq.\,(\ref{def_Gamma})
should be regarded as a function of $\phi$ alone,
we eliminate $J$ 
using Eq.\,(\ref{def_Sym}) to obtain
\begin{align}
 \Gamma[\phi] &= -W\left[\frac{\delta \Gamma}{\delta \phi}[\phi]
 \right] + \frac{\delta \Gamma}{\delta \phi}[\phi] \phi\\
 &= -\log{\left\{
 \int \mathrm{d}\chi \mathrm{e}^{-S[\chi]+\frac{\delta \Gamma}{\delta
 \phi}[\phi]\chi}\right\} }
 + \frac{\delta \Gamma}{\delta \phi}[\phi]\phi\\
 &= -\log{\left\{
 \int \mathrm{d}\chi \mathrm{e}^{-S[\chi]+\frac{\delta \Gamma}{\delta
 \phi}[\phi](\chi-\phi)}
\right\} }. \label{calc_ea}
\end{align}
This suggests a means of calculating the effective action.
However, it requires a recursive procedure
that includes integrations over all possible combinations of control
variables $\chi$, which appears to be intractable.

To compute the effective action,
we may evaluate this integral stepwise, 
using methods developed in renormalization group theory \cite{zee2010quantum}.
A relevant concept that we shall explore later is
the effective average action $\Gamma_k[\phi]$ proposed by
Wetterich \cite{Wetterich199390},
which constitutes a one-parameter family of functionals
interpolating between the action $S[\phi]$ and the effective action
$\Gamma[\phi]$.

The aim of this paper is to propose a possible
framework that will help solve the multiscale data assimilation problem
by replacing the cost function with the effective average action.
We also propose a novel method for evaluating the gradient of the effective average action numerically.
In principle, this enables us to solve a broader range of data
assimilation problems
by seeking the stationary point of the effective average action 
using its gradient,
which is referred to as
 the effective gradient or the coarse-grained sensitivity.

The concept of the effective average action is explained in
section \ref{sec_def},
and the meaning of its stationary point is clarified in section \ref{sec_prop}.
Section \ref{sec_calc} describes a novel procedure for calculating the
gradient of the effective action.
Sections \ref{sec_ex1} and  \ref{sec_ex2} illustrate some
applications of the method to data assimilation or sensitivity studies.

\section{Effective average action}
When dealing with a multiscale system,
it is often difficult to define  the sensitivity with respect to
the control variable, because
fast degrees of freedom may have many statistical paths
related to the sensitivity that cannot
 be expressed in a deterministic manner.
 In other words, we cannot always
 use the sensitivity to choose the optimal realization of
 a fluctuation from among
 a large ensemble of fast fluctuations in the control space.
This motivates the definition of a macroscopic field in which fast degrees of
freedom are treated as averaged quantities.
A suitable  tool for this purpose
is the effective average action \cite{Wetterich199390}.

   \subsection{Definition}\label{sec_def}

The procedure for the effective action reviewed in the Introduction,
$S[\chi] \to W[J] \to \Gamma[\phi]$,
can also be applied to the derivation of the effective average action.
This introduces some filtering terms,
$\Delta S_k[\chi]$ to Eq.\,(\ref{def_Z}) and $-\Delta S_k[\phi]$ to
Eq.\,(\ref{def_Gamma}),
which have the  effect of selectively integrating out 
the fast degrees of freedom in the control space
to enable  the dynamics of slower variables to be investigated.

We start by defining an infra-red filter\footnote{
We use the following matrix-form notation to denote the spatio-temporal
(or momentum space) integration of fields:
\begin{align*}
 \int d^dx  \sum_n J_n(x) \chi_n(x)
 & \to
J^T \chi,\\
  \frac12 \int \frac{d^dq}{(2\pi)^d} \sum_{m,n} \chi_m(q) \left( R_k
 \right)_{m,n}(q)\chi_n(-q)
& \to
 \frac12 \chi^T R_k \chi,\\
 \frac{\delta S}{\delta \phi(x)}[\phi]
& \to
 \frac{\delta S}{\delta \phi}[\phi],
\end{align*}
where $d$ is the dimension of the space-time,
and $m$, $n$ are the indices of the components.},
  \begin{align}
   \Delta S_k[\chi] \equiv \frac12 \chi^T R_k \chi,
  \end{align}
where $\chi \in \mathbb{R}^M$ is the control variable in a discrete space-time with
$M$ cells, and $R_k  \in \mathbb{R}^M \times \mathbb{R}^M$ is a discrete low-pass filter.

To derive concrete expressions for $R_k$,
let us consider a simple case with a
cyclic control variable $\chi$
in a 1-dimensional discrete domain $l=1,2,\cdots,M$.
We  define the discrete Fourier transform $\widehat{\chi}$ of $\chi$
and its inverse as:
\begin{align}
\widehat{\chi}_j &= \frac1{\sqrt{M}} \sum_{l=1}^{M} \chi_l
 \mathrm{e}^{-\frac{2\pi j l}{M}\mathrm{i}},\quad |j|<[M/2],\\
\chi_l &= \frac1{\sqrt{M}} \sum_{j=-[M/2]}^{[M/2]} \widehat{\chi}_j
 \mathrm{e}^{\frac{2\pi j l}{M}\mathrm{i}},\quad l=1,2,\cdots,M,
\end{align}
where $M$ is odd for simplicity,
$[\cdot]$ denotes the roundoff, and
$\mathrm{i}$ is the imaginary unit.
If we assume the infra-red filter is represented by a cutoff
of high-wavenumber modes, then:
 \begin{align}
  \Delta S_k[\widehat{\chi}]
  &= \frac12 \sum_{j} \widehat{\chi}_{-j} \widehat{R_k}(j) \widehat{\chi}_{j},\\
\widehat{R_k}(j) &=
  \begin{cases}
   k^2\left( 1- \frac{j^2}{j_k^2} \right) & \text{if} \;
   |j| < j_k
   \\
  0 & \text{otherwise}\label{ex_Rk}
 \end{cases},
 \end{align}
where $j,j_k \in \mathbb{Z}$, and $j_k$ is the cutoff level.
The filtering term can then be written as
\begin{align}
 \Delta S_k[\widehat{\chi}] &=
 \frac12 \sum_{|j| < j_k}
 \widehat{\chi}_{-j} k^2 \left( 1- \frac{j^2}{j_k^2} \right)
 \widehat{\chi}_{j}, \nonumber\\
&=\frac12 \sum_{l,l'=1}^{M} \chi_{l} \left\{
 \sum_{|j| < j_k}  \frac{k^2}{M} \left( 1- \frac{j^2}{j_k^2} \right)
 \mathrm{e}^{\frac{-2\pi j
 (l-l')}{M}\mathrm{i}}
\right\}\chi_{l'} \\
&=\frac12 \sum_{l,l'=1}^{M} \chi_{l} \left\{
 \sum_{|j| < j_k}
 \frac{k^2}{M} \left( 1- \frac{j^2}{j_k^2} \right)
 \cos{\left[
\frac{2\pi j
 (l-l')}{M}\right]}
\right\}\chi_{l'}.
\end{align}
The expression in the curly brackets gives a matrix representation of $R_k$
in position space.
If $M$ is even, we can replace Eq.\,(\ref{ex_Rk}) by
\begin{align}
\widehat{R_k}\left(j+\frac12\right) &=
  \begin{cases}
   k^2\left( 1- \frac{\left(j+\frac12\right)^2}{j_k^2} \right) & \text{if} \;
   |j+\frac12| < j_k
   \\
  0 & \text{otherwise}\label{ex_Rk_even}
 \end{cases}
 \end{align}
where $j,j_k \in \mathbb{Z}$.

To further simplify the filter, we can also use
$R_k=k^2$ with $j_k \to \infty$, which yields
\begin{align}
 \Delta S_k[\chi]
&=
\frac{k^2}{2}
 \sum_{l=1}^{M} \chi_{l}^2. 
\end{align}

As we will see later,
the filter should have the following properties:
\begin{align}
k \to \infty &\Rightarrow  R_k \to \infty,\label{prop1_Rk}\\
k \to 0 &\Rightarrow  R_k \to 0. \label{prop2_Rk}
\end{align}

With the filtering term, the partition function $Z_k$ and its logarithm
  for the high-wavenumber modes
  can be defined as:
  \begin{align}
   Z_k[J] &= \int \mathrm{d}\chi
   \mathrm{e}^{-S[\chi]-\Delta S_k[\chi]+J^T\chi},\label{def_Zk}\\
   W_k[J] &= \log{Z_k[J]},
  \end{align}
  where $J$ is the external field.
As $\Delta S_k[\chi]$ is large for low-wavenumber modes,
  the term $\exp{\left(-\Delta S_k[\chi]\right)}$
has the effect of focusing
the integration  on
the high-wavenumber modes in $\chi$.

Applying the Legendre transformation to switch the independent variable
$J$ to
$\phi$, we obtain the effective average
action \cite{Wetterich199390}:
\begin{align}
\Gamma_k[\phi] &\equiv \sup_J{\left\{-W_k[J] + J^T
 \phi \right\} } - \Delta S_k[\phi].\label{Legendre_Gammak}
\end{align}
Note that
this transform should have
the additional term $- \Delta S_k[\phi]$ for the following reason \cite{Wetterich199390}.
From the supremum condition, we find:
  \begin{align}
   \phi &= \left(\frac{\delta W_k}{\delta J}[J]\right)^T
= \left< \chi \right>_{k,J}
\equiv
\frac{ \int \mathrm{d}\chi~\chi~\mathrm{e}^{-S[\chi]-\Delta S_k[\chi]+J^T\chi}}
     { \int \mathrm{d}\chi      \mathrm{e}^{-S[\chi]-\Delta
   S_k[\chi]+J^T\chi}} \label{def_ave}\\
&=
\frac{ \int \mathrm{d}\chi~(\phi+\chi)~\mathrm{e}^{-S[\phi+\chi]-\Delta
 S_k[\chi]+(J^T-\phi^T R_k)\chi}}
     { \int \mathrm{d}\chi \mathrm{e}^{-S[\phi+\chi]-\Delta
 S_k[\chi]+(J^T-\phi^T R_k)\chi}}.\label{def_phi}
  \end{align}
  Eq.\,(\ref{def_ave}) appears to indicate
  that $\phi$ is analogous to the conditional
  expectation of the control variable
  in the vicinity of $\chi=0$
  under the existence of the
  external field $J$.
  However, if we insert
  into Eq.\,(\ref{def_phi})
  the relation:
  \begin{align}
   \frac{\delta \Gamma_k}{\delta \phi}[\phi]&=
   J^T-\frac{\delta \Delta S_k}{\delta \phi}[\phi],\label{dgdp}
  \end{align}
  which is derived by the same operation as in Eq.\,(\ref{derive_J}),
  we obtain
\begin{align}
 \phi
&=
\frac{ \int \mathrm{d}\chi~(\phi+\chi)~\mathrm{e}^{-S[\phi+\chi]-\Delta
 S_k[\chi]+\frac{\delta \Gamma_k}{\delta \phi}[\phi]\chi}}
     { \int \mathrm{d}\chi \mathrm{e}^{-S[\phi+\chi]-\Delta
 S_k[\chi]+\frac{\delta \Gamma_k}{\delta \phi}[\phi]\chi}}.\label{def_phi2}
  \end{align}
This shows that
  $\phi$ is  in fact analogous to the conditional
  expectation of the control variable
  in the vicinity of $\phi$ itself
  under the existence of the
  external field $\delta \Gamma_k[\phi]/\delta \phi$.
  Thus, the term $-\Delta S_k[\phi]$ in Eq.\,(\ref{Legendre_Gammak}) ensures
  that 
  $\phi$ is always the average of the surrounding $\chi$.

 From Eq.\,(\ref{def_ave}) and
 the derivative of Eq.\,(\ref{dgdp}) with respect to $\phi$,
we can
 see that
  the effective average action $\Gamma_k[\phi]$ also satisfies the
  following equality:
\begin{align}
\frac{\delta^2 \Gamma_k}{\delta \phi^2}[\phi]
&= \frac{\delta J}{\delta \phi}-\frac{\delta^2 \Delta S_k}{\delta
 \phi^2}\\
&=
\left(\frac{\delta^2 W_k}{\delta J^2}[J]\right)^{-1}-\frac{\delta^2 \Delta S_k}{\delta
 \phi^2}\\
&=
\left( \left< \chi \chi^T \right>_{k,J} - \phi
\phi^T \right)^{-1}
-\frac{\delta^2 \Delta S_k}{\delta \phi^2}.\label{d2gdp2}
\end{align}

Eliminating $J$ from Eq.\,(\ref{Legendre_Gammak}) 
using Eq.\,(\ref{dgdp}), we have
\begin{align}
 \Gamma_k[\phi] &= -W\left[\frac{\delta \Gamma_k}{\delta \phi}[\phi]
+\frac{\delta \Delta S_k}{\delta \phi}[\phi]
 \right] +
\left(
\frac{\delta \Gamma_k}{\delta \phi}[\phi]
+\frac{\delta \Delta S_k}{\delta \phi}[\phi]
\right)
 \phi
- \Delta S_k[\phi]
 \\
 &= -\log{\left\{
 \int \mathrm{d}\chi \mathrm{e}^{-S[\chi]+\frac{\delta \Gamma_k}{\delta
 \phi}[\phi](\chi-\phi) -\Delta S_k[\chi-\phi]}
\right\} }\\
&= -\log{\left\{\int \mathrm{d}\chi \mathrm{e}^{-S[\phi+\chi]
 +\frac{\delta \Gamma_k}{\delta \phi}[\phi]\chi-\Delta S_k[\chi]}
 \right\} }.\label{gamma1}
\end{align}
Although this has a recursive form about $\Gamma_k[\phi]$,
we can write an approximation in closed form (see Appendix \ref{app1} for the derivation):
\begin{align}
\Gamma_k[\phi]
&\simeq S[\phi]+\frac12 \log{\det{\left\{\frac{\delta^2 }{\delta
 \phi^2}\left(S+\Delta S_k\right)[\phi] \right\}} }.
\end{align}
Note that we will not resort to such
perturbation expansions in our numerical calculation,
because it requires
higher derivatives of the action, which are
 not always easy to calculate.

\begin{figure}
\begin{center}
\includegraphics[width=36em]{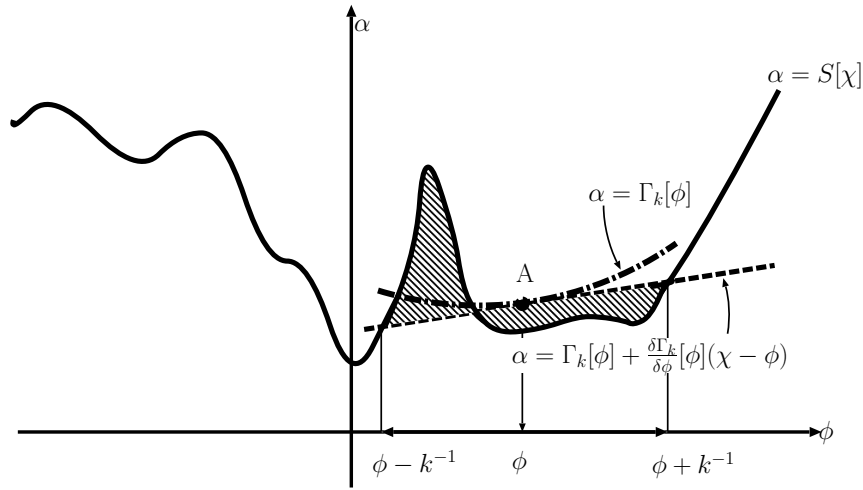}
\end{center}
 \caption{
 The concept of the effective average action $\Gamma_k$
 (dotted-dashed curve).
The tangent point $A(\phi,\Gamma_k[\phi])$ and
the slope $\delta \Gamma_k[\phi]/\delta \phi$
are such that
the point $\phi$ coincides with
the weighted average of the points $\chi$
around the interval $\phi-k^{-1} \leq \chi \leq \phi+k^{-1}$.
The weight exponentiates
 the deviation (shaded region) of the action $\alpha=S[\chi]$ (solid curve)
 from
 the tangent plane $\alpha = \Gamma_k[\phi]+
 \left(\delta \Gamma_k[\phi]/\delta \phi \right)(\chi-\phi)$ (dashed line).
For simplicity, it is assumed that $\chi$ is one-dimensional and
the filter has $R_k=k^2$.
}
\label{fig_phi}
\end{figure}

Fig.\,\ref{fig_phi} illustrates the relation between
the action and the effective average action in a simple case,
where we assume $\chi \in \mathbb{R}^1$ and
the filter is $R_k=k^2$.
The tangent point $A(\phi,\Gamma_k[\phi])$ and
the slope $\delta \Gamma_k[\phi]/\delta \phi$
are such that
the point $\phi$ coincides with
the weighted average of the points $\chi$
around the interval $\phi-k^{-1} \leq \chi \leq \phi+k^{-1}$.
Indeed, we see from Eqs.\,(\ref{def_phi2}) 
 and (\ref{gamma1}) that
\begin{align}
\phi &=
\int \mathrm{d}\chi~ \chi ~
 \mathrm{e}^{-S[\chi]+\Gamma_k[\phi]
+\frac{\delta \Gamma_k}{\delta \phi}[\phi]
 (\chi-\phi) -\frac{k^2}{2} \left( \chi-\phi \right)^2}.
\end{align}
One interesting thing about this smoothing is that
it is not an averaging of the value of $S[\chi]$ but of its independent
variable $\chi$.
Thereby, the effective average action serves
as a kind of smoothed version of the cost function.

Taking Eqs.\,(\ref{prop1_Rk}) and (\ref{prop2_Rk}) into account,
$k\to \infty$ implies
$
\mathrm{e}^{-\Delta S_k[\chi]} =
\mathrm{e}^{-\frac12 \chi^T R_k \chi}
\to \delta[\chi]$, which leads Eq.\,(\ref{gamma1})
to
\begin{align}
 \Gamma_{k \to \infty}[\phi] = S[\phi].
\end{align}
This means that when $k$ is sufficiently large
no fields around $\phi$ are counted in
$\Gamma_k[\phi]$, other than the field $\phi$ itself,
and thus
the effective average action approaches the original action, or the
cost function.
In contrast, it is apparent that
  \begin{align}
   \Gamma_{k \to 0}[\phi]&=\Gamma[\phi].
   \end{align}
Hence, we have confirmed that
$\Gamma_k[\phi]$
constitutes a one-parameter family of functionals
interpolating between the action $S[\phi]$ and the effective action
$\Gamma[\phi]$.

Hereafter, we assume that $k$ has a finite value so as to integrate out some modes.

\subsection{Property of the stationary point}\label{sec_prop}
We assume that the stationary problem:
\begin{align}
\frac{\delta \Gamma_k}{\delta \phi}[\phi]
&= J^T-\frac{\delta \Delta S_k}{\delta \phi}[\phi] = 0\label{motion}
\end{align}
has stationary values at $\hat{\phi}$.
From definition (\ref{def_ave}), we have:
\begin{align}
\hat{\phi} &=
\left< \chi \right>_{k,\frac{\delta \Delta S_k}{\delta
 \phi}[\hat{\phi}]}.
\label{stationarity}
\end{align}
Since Eq.\,(\ref{motion}) can be thought of as the statistical equation
of motion for the field $\phi$, the solution $\hat{\phi}$ offers
the estimated path for the statistical problem.
At the stationary point, Eq.\,(\ref{def_phi}) reads:
\begin{align}
\hat{\phi} &= \frac
{\int \mathrm{d}\chi ~ (\hat{\phi}+\chi) ~
 \mathrm{e}^{-S[\hat{\phi}+\chi]-\Delta S_k[\chi] }}
{\int \mathrm{d}\chi ~
 \mathrm{e}^{-S[\hat{\phi}+\chi]-\Delta S_k[\chi] }}.
\end{align}
This means that the stationary value $\hat{\phi}$ provides the  average
with respect to high-wavenumber modes (see also Appendix \ref{app2}).

\section{Estimation of the effective gradient}\label{sec_calc}
\subsection{Definition as an expected value}
Rewriting Eq.\,(\ref{gamma1}), we introduce the
exponent $\mathfrak{R}[\phi,\chi]$ for convenience:
\begin{align}
\mathrm{e}^{-\Gamma_k[\phi]}
&=
\mathrm{e}^{-S[\phi]}
\int \mathrm{d} \chi
\mathrm{e}^{-\mathfrak{R}[\phi,\chi]},\\
\mathfrak{R}[\phi,\chi]&\equiv
S[\phi+\chi]-S[\phi]-\frac{\delta \Gamma_k}{\delta \phi}[\phi] \chi
 +\Delta S_k[\chi],\label{defR}\\
\Gamma_k[\phi] &= S[\phi] -\log{
\int \mathrm{d}\chi \mathrm{e}^{-\mathfrak{R}[\phi,\chi]}}.\label{defGamma3}
\end{align}
The gradient of $\Gamma_k[\phi]$ is derived as the expected value $\left< \cdot \right>_{\mathfrak{R}}$ under
the weight $\mathrm{e}^{-\mathfrak{R}}$:
\begin{align}
 \frac{\delta \Gamma_k}{\delta \phi}[\phi] &=
 \frac{\delta S}{\delta \phi}[\phi]
-\frac{
\int \mathrm{d}\chi \left(-\frac{\delta \mathfrak{R}}{\delta \phi}[\phi,\chi]\right)
\mathrm{e}^{-\mathfrak{R}[\phi,\chi]}
}{
\int \mathrm{d}\chi \mathrm{e}^{-\mathfrak{R}[\phi,\chi]}
}\nonumber\\
&=
\left<
\frac{\delta S}{\delta \phi}[\phi+\chi]
\right>_{\mathfrak{R}}
-  \left< \chi \right>_{\mathfrak{R}}^T
\frac{\delta^2 \Gamma_k}{\delta \phi^2}[\phi].\label{eaa_en0}
\end{align}
From Eq.\,(\ref{def_phi2}), we have that
\begin{align}
 0&=\int \mathrm{d}\chi ~ \chi~ \mathrm{e}^{-\mathfrak{R}[\phi,\chi]} \propto \left< \chi \right>_{\mathfrak{R}}.
\end{align}
Thus, Eq.\,(\ref{eaa_en0}) can be simplified to
\begin{align}
 \frac{\delta \Gamma_k}{\delta \phi}[\phi] &=
\left<
\frac{\delta S}{\delta \phi}[\phi+\chi]
\right>_{\mathfrak{R}}.\label{eaa_en}
\end{align}

Eqs.\,(\ref{defR}) and (\ref{eaa_en})
are recursive with respect to $\delta \Gamma_k[\phi]/\delta \phi$.
Therefore, we need some approximation
 to enable a numerical evaluation.
We may replace $\delta \Gamma_k[\phi]/\delta \phi$
with the approximation $\delta S[\phi]/\delta \phi$ in $\mathfrak{R}[\phi,\chi]$
on the right-hand side of the equation.
Eq.\,(\ref{eaa_en}) can then be evaluated
  using the Metropolis method
\cite{metropolis1953equation,hastings1970monte}.
We can then apply a successive correction procedure
by updating the expectation 
with the latest value of
$\delta \Gamma_k[\phi]/\delta \phi$
in the weight.

Furthermore,
from Eqs.\,(\ref{def_ave}), (\ref{def_phi2}),
and (\ref{d2gdp2}),
we find that the second derivative can also be derived as
the expected value:
\begin{align}
\frac{\delta^2 \Gamma_k}{\delta \phi^2}[\phi] &=
\left< \chi \chi^T \right>_{\mathfrak{R}}^{-1} - \frac{\delta^2 \Delta S_k}{\delta \phi^2}.
\end{align}

\subsection{Evaluation through the Metropolis method}
In contrast to the case of the effective action
 in Eq.\,(\ref{calc_ea}),
the filtering term $\Delta S_k[\chi]$
in Eq.\,(\ref{defR}) has the effect of
confining the weight $\exp{(-\mathfrak{R}[\phi,\chi])}$
to a small region in the control space, as illustrated
in Fig.\,\ref{fig_phi}. Owing to this,
we may assume that the expected value will be
efficiently evaluated by a Markov-chain Monte Carlo method,
e.g., the Metropolis adjusted Langevin algorithm \cite{roberts1996exponential}.
Using the fact that the Langevin equation:
\begin{align}
 d\chi_t &= \frac12 \nabla \log{f(\chi_t)} dt + dW_t, \quad (W_t:\text{the Wiener process})
\end{align}
has the invariant distribution $\pi(\chi) \equiv f(\chi)/\int \mathrm{d}\chi f(\chi)$,
 we construct a Markov chain by discretizing the equation and applying
 an acceptance/rejection procedure.

At time step $n$, according to the weight:
\begin{align}
 f(\chi^{(n)}) &= \mathrm{e}^{-\mathfrak{R}[\phi,\chi^{(n)}]},
\end{align}
we define a proposal normal distribution:
\begin{align}
 q\left(\chi|\chi^{(n)}\right) &=
 \mathcal{N}\left(\chi^{(n)}-\frac{\sigma^2}{2}\nabla_{\chi^{(n)}}
 \mathfrak{R}[\phi,\chi^{(n)}], \sigma^2 I \right),\\
\nabla_{\chi^{(n)}}\mathfrak{R}[\phi,\chi^{(n)}]
 &=
 \left( \frac{\delta \mathfrak{R}}{\delta \chi^{(n)}}[\phi,\chi^{(n)}]\right)^T,\\
 \frac{\delta \mathfrak{R}}{\delta \chi^{(n)}}[\phi,\chi^{(n)}]
 &=
 \frac{\delta S}{\delta \phi}[\phi+\chi^{(n)}]
-
 \frac{\delta \Gamma_k}{\delta \phi}[\phi]
 +
 \frac{\delta \Delta S_k}{\delta \phi}[\chi^{(n)}]\\
 &\simeq
 \frac{\delta S}{\delta \phi}[\phi+\chi^{(n)}]
-
 \frac{\delta S}{\delta \phi}[\phi]
 +
 \frac{\delta \Delta S_k}{\delta \phi}[\chi^{(n)}]\label{RinMALA}
\end{align}
to generate a random field $\chi^*$ that obeys $q$:
\begin{align}
 \chi^* &= \chi^{(n)} - \frac{\sigma^2}{2}\nabla_{\chi^{(n)}} \mathfrak{R}[\phi,\chi^{(n)}]
 + \sigma \xi, \quad \xi \sim \mathcal{N}(0,I).
\end{align}
We then update $\chi^{(n+1)}=\chi^*$
with the acceptance probability:
\begin{align}
\rho\left( \chi^{(n)},\chi^*\right) &=\min{\left(1,
 \frac{f(\chi^*)}{f(\chi^{(n)})}
 \frac{q(\chi^{(n)}|\chi^*)}{q(\chi^*|\chi^{(n)})}
\right)},
\end{align}
or retain $\chi^{(n+1)}=\chi^{(n)}$.
The ensemble of sample sequences $\chi^{(n)}$ drawn in this way
approximately follows the invariant distribution $\pi(\chi)$,
and  we can estimate the expected value accordingly:
\begin{align}
\left< \frac{\delta S}{\delta \phi}[\phi+\chi] \right>_{\mathfrak{R}}
&\simeq
\frac1N \sum_{n=1}^{N} \frac{\delta S}{\delta \phi}[\phi+\chi^{(n)}].
\label{Metro}
\end{align}
Note that, after the averaging,
we may perform another refined averaging by
substituting the derived sensitivity into the weight,
because the term
$\left(\delta S[\phi]/\delta\phi\right)^T$
in Eq.\,(\ref{RinMALA}) should have been
$\left(\delta \Gamma_k[\phi]/\delta\phi\right)^T$.

Furthermore,
within the limit of the accuracy of importance sampling,
the finite difference of
$\Gamma_k$ can also be estimated as
\begin{align}
\Gamma_k[\phi+\Delta \phi]-\Gamma_k[\phi]
&=
-\log \left<
\mathrm{e}^{-\mathfrak{R}[\phi+\Delta \phi,\chi]
+\mathfrak{R}[\phi,\chi]}
\right>_{\mathfrak{R}}.
\end{align}

\section{A simple example}\label{sec_ex1}
\begin{figure}
\begin{center}
\includegraphics[width=36em]{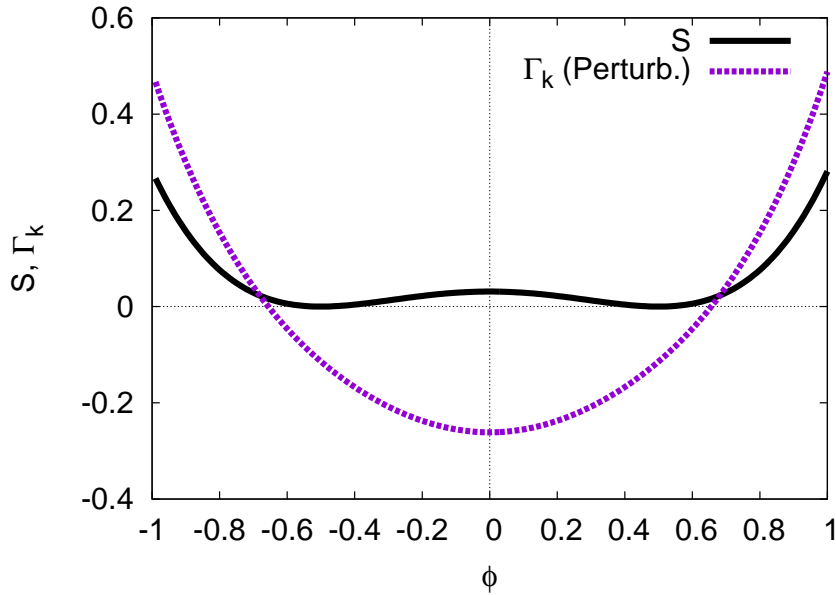}
\end{center}
 \caption{Action $S[\phi]$ (black) and effective average action $\Gamma_k[\phi]$
 evaluated using a perturbation expansion (purple) for the double-well potential}
\label{fig_doublewell}
\end{figure}

\begin{figure}
\begin{center}
\includegraphics[width=36em]{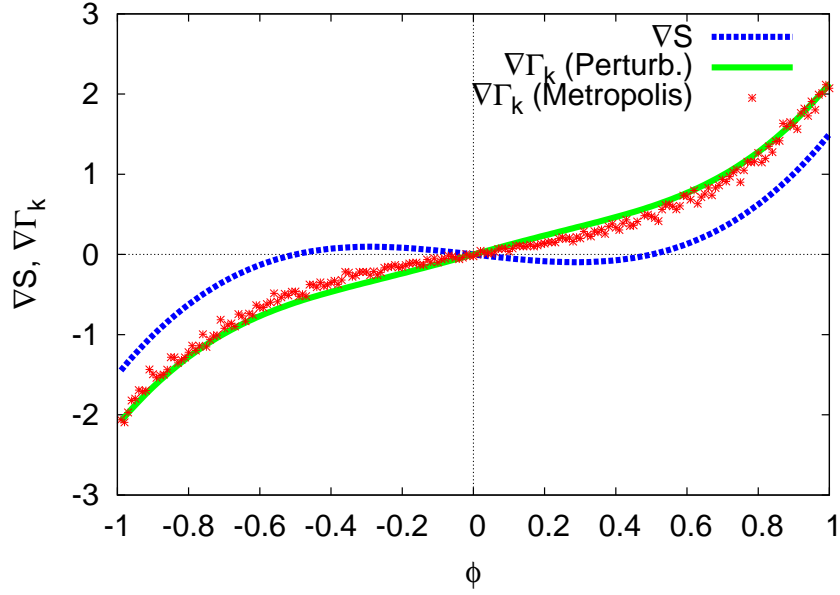}
\end{center}
 \caption{Gradient of action $\delta S[\phi]/\delta \phi$ (blue),
effective gradient of action $\delta \Gamma_k[\phi]/\delta \phi$ 
evaluated using a perturbation expansion (green),
 and effective gradient of action
 evaluated by the Metropolis method
 (red)  for the double-well potential}
\label{fig_doublewell2}
\end{figure}

As a simple example, we consider a double-well potential:
\begin{align}
S[\phi]&=\frac12 \left( \phi^2 -a^2 \right)^2,\quad
 a=0.5,  \quad R_k = 2^2.
\end{align}
The effective average action given by a perturbation expansion up to the second
order (see Appendix \ref{app1}) is:
\begin{align}
 \Gamma[\phi]&=S[\phi]+\frac12 \log{( 6\phi^2-2a^2+R_k )}.
 \end{align}
Fig.\,\ref{fig_doublewell} shows
that the potential barrier at the center (black)
is eliminated (purple) by integrating out the fluctuation.
The gradients, along with the effective gradient of the action evaluated
by the Metropolis method, are shown in Fig.\,\ref{fig_doublewell2}.
These curves suggest that
the variational method using the coarse-grained sensitivity
can capture the expected value $\hat{\phi}=0$,
which
traditional variational methods
 will fail to find.
This is because, in principle,
variational methods are all designed to
find one of the stationary values of the cost function,
in this case $\phi=\pm a$.

\section{Application to data assimilation}\label{sec_ex2}
\subsection{Quadratic cost function}
The cost function used in data assimilation is usually a quadratic
in the nonlinear functional $F[\phi]$, such as\footnote{
In particular, strong-constraint 4D-Var has
\begin{align*}
 F[\phi] &=
  \begin{bmatrix}
   B^{-\frac12}\left( \phi - m_0 \right) \\
   R^{-\frac12}\left( y - \Psi\left[ \phi \right] \right)
  \end{bmatrix},
\end{align*}
where
$\Psi$ is the model,
$B$, $R$ are the background and observational error covariance
matrices, respectively,
and $m_0$,  $y$ are the first-guess field and observational data, respectively.
}:
\begin{align}
  S[\phi]\equiv\frac12 F[\phi]^T F[\phi],
  \quad \frac{\delta S}{\delta \phi}[\phi]=
F[\phi]^T \frac{\delta F}{\delta \phi}[\phi].
\end{align}
To obtain the averaged sensitivity
$
\delta \Gamma_k[\phi]/\delta \phi
=
\left< \delta S[\phi+\chi]/\delta \phi \right>_{\mathfrak{R}}$,
we perform the following calculations:
\begin{align}
\left(\frac{\delta S}{\delta \phi}[\phi+\chi]\right)^T&=
\left(\frac{\delta F}{\delta \phi}[\phi+\chi]\right)^T F[\phi+\chi],\\
 \mathfrak{R}[\phi,\chi] &=
\frac12 F[\phi+\chi]^T F[\phi+\chi]
- \frac12 F[\phi]^T F[\phi]\nonumber\\
&- \chi^T \left(
\frac{\delta \Gamma_k}{\delta \phi}[\phi]
\right)^T
+\Delta S_k[\chi]\\
&\simeq
\frac12 F[\phi+\chi]^T F[\phi+\chi]
- \frac12 F[\phi]^T F[\phi]\nonumber\\
&- \chi^T \left(\frac{\delta F}{\delta \phi}[\phi]\right)^T F[\phi]
+\Delta S_k[\chi], \label{RinGamma}\\
\nabla_{\chi} \mathfrak{R}[\phi,\chi] &=
\left(\frac{\delta F}{\delta \phi}[\phi+\chi]\right)^T F[\phi+\chi]
-
 \left(
 \frac{\delta \Gamma_k}{\delta \phi}[\phi]
 \right)^T\nonumber\\
&+\left( \frac{\delta \Delta S_k}{\delta \chi}[\chi]\right)^T.
\end{align}
Thus, each sample requires a forward integration
$F[\phi+\chi]$ and a subsequent adjoint integration
$\left(\delta F[\phi+\chi]/\delta \phi \right)^T$.

\subsection{The Logistic map}\label{Logi}
We now consider a smoothing problem for the Logistic map
$\Psi \mapsto r \Psi (1-\Psi)$ \cite{may1976simple,LSZ15} fitted to observation $y$.
As a data assimilation problem, we use the following cost function $S$
and its gradient\footnote{
In this case,
$F[\phi] =
  \begin{bmatrix}
   (\phi - m_0)/\sigma_0, &
   (y_1  - \Psi^{(1)}[\phi])/\gamma,  &   \cdots, &
   (y_J  - \Psi^{(J)}[\phi])/\gamma
  \end{bmatrix}^T$.}:
\begin{align}
 S[\phi] &= \frac{1}{2\sigma_0^2} (\phi-m_0)^2 +
\sum_{j=0}^{J-1} \frac{1}{2\gamma^2}\left( y_{j+1} -
 \Psi^{(j+1)}[\phi]\right)^2, \label{cost_logi}
\end{align}
\begin{align}
\Psi^{(j+1)}[\phi] &= r \Psi^{(j)}[\phi]
 \left(1-\Psi^{(j)}[\phi]\right),
\quad \Psi^{(0)}[\phi] = \phi, \\
\frac{\delta S}{\delta \phi}[\phi]&=
\frac{1}{\sigma_0^2}(\phi-m_0)
+\frac{1}{\gamma^2} \sum_{j=0}^{J-1}
 \left\{\prod_{l=0}^j
r(1-2\Psi^{(l)}[\phi])
\right\}\nonumber\\
&\times \left(\Psi^{(j+1)}[\phi]-y_{j+1}\right),
\end{align}
where $\sigma_0^2$
and  $\gamma^2$ are the
background and observational
error variances, respectively.
The parameters are set to $r=4$, $J=6$, $\sigma_0=0.1$, $\gamma=0.2$, and
$m_0=0.4$ (first-guess), similar to those in \cite{LSZ15}.
The observations are sampled from a model sequence given by the initial
value $v_0=0.3$ added to observational noise.

\begin{figure}
\begin{center}
 \includegraphics[width=36em]{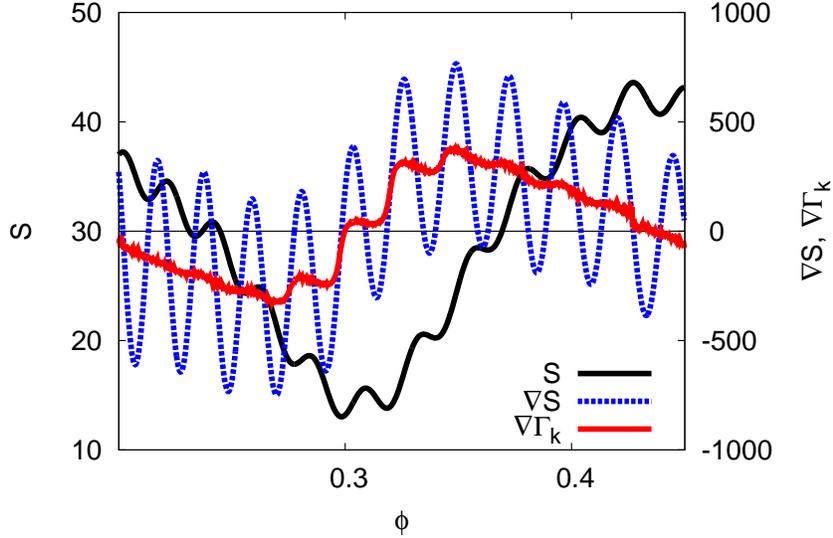}
\end{center}
\caption{Action $S[\phi]$ (black), 
gradient of action $\delta S[\phi]/\delta \phi$
 (blue), and effective gradient of action $\delta \Gamma_k[\phi]/\delta \phi$ 
 (red) for the Logistic map. The true value for the data
 assimilation problem is
 $\phi=0.3$, and the first-guess is $0.4$.}
\label{fig_logistic}
\end{figure}

The optimization  problem
can apparently be solved  using a variational method
that seeks the optimal initial condition using the gradient information.
However, there are multiple extrema of the cost function (see black curve
in Fig. \ref{fig_logistic}), which makes it difficult to find the global minimum.
Thus, we should utilize the effective gradient
(\ref{Metro}), derived
 using the Metropolis method.
We use a finite constant $R_k=(0.008)^{-2}$,
where $0.008$ is the typical half-wavelength of short fluctuations in
$S[\phi]$.

Fig.\,\ref{fig_logistic} shows the
action (black), the gradient of action (blue), and the effective
 gradient of action (red) for this system.
It is clear that the original gradient has too many zeros for
worthwhile  variational data assimilation using the
gradient.
However, because  the effective gradient has relatively few zeros,
it can be applied to
 a variational data assimilation to find the
minimum at around $\phi=v_0$, as long as the first-guess is not far
 from the true value.

We performed  two data assimilation experiments using the steepest descent method
with the gradient $\delta S[\phi]/\delta \phi$
 and the effective gradient $\delta \Gamma_k[\phi]/\delta \phi$
 (see the algorithm in Appendix \ref{app3}).
As shown in Fig.\,\ref{fig_cost},
the case with $\delta S[\phi]/\delta \phi$ converges to a local minimum $\phi=0.388$,
whereas the case with $\delta \Gamma_k[\phi]/\delta \phi$
 converges to $\phi=0.315$ around the global minimum,
which indicates the superiority of the effective gradient.
Note that the decrease in the cost function $S[\phi]$ in the latter case
is not necessarily monotonic, because
the optimization problem is actually
defined for the effective average action $\Gamma_k[\phi]$.

This illustrates the potential usefulness of the effective gradient (sensitivity)
for data assimilation.

\begin{figure}
\begin{center}
 \includegraphics[width=30em]{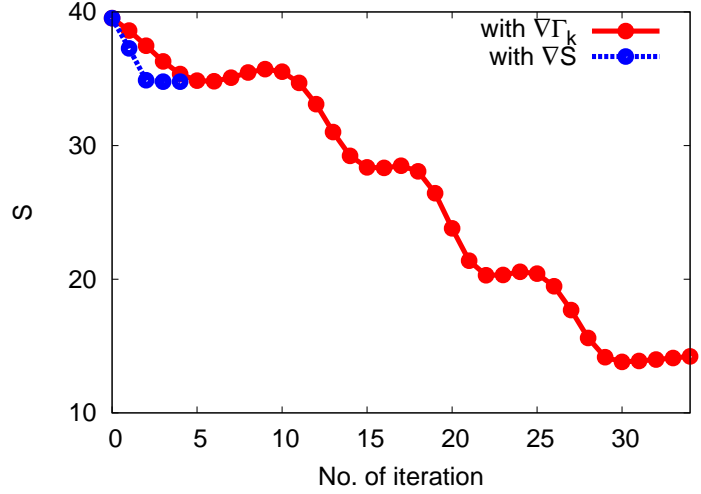}
\end{center}
\caption{The variation of the cost function during data assimilation using the
 steepest descent method with $\delta S[\phi]/\delta \phi$ (blue)
 and  $\delta \Gamma_k[\phi]/\delta \phi$
 (red).
The former converges to a local minimum $\phi=0.388$; the latter
 converges to $\phi=0.315$, which is close to the global minimum.}
\label{fig_cost}
\end{figure}

\subsection{Lorenz '96 model}\label{L96_1}
Next, we examine the sensitivity that appears
during data assimilation in the Lorenz '96 model,
which is a simple dynamical system
designed to mimic the dynamics of Rossby waves in
atmospheric dynamics \cite{l96,LSZ15}.
This system can be written as:
\begin{align}
 \frac{d\theta_l}{dt} &= \theta_{l-1} \left( \theta_{l+1} - \theta_{l-2} \right) -\theta_l
 +F,\quad l=1,2,\cdots, M, \label{l96model}\\
\theta_0 &= \theta_M, \quad \theta_{M+1} = \theta_1, \quad \theta_{-1} = \theta_{M-1}.
\end{align}
We write the time evolution operator from the initial condition
 as:
\begin{align}
\Psi^{(j)}\left[ \theta(t=0) \right]&=\theta(t=t_j),
\end{align}
with time-step $t_{j+1}-t_j=\Delta t$.
If we choose
the initial condition $\theta(t=0) = \phi \in \mathbb{R}^M$ as
the control variable,
we can define a similar cost function to that in Eq.\,(\ref{cost_logi}):
\begin{align}
 S[\phi] &= \frac{1}{2\sigma_0^2} \|\phi-m_0\|^2 +
\sum_{j=0}^{J-1} \frac{1}{2\gamma^2}\left\|y_{j+1} -
 \Psi^{(j+1)}[\phi]\right\|^2, \label{cost_l96}
\end{align}
where $\|\cdot\|$ is the Euclidean norm.
The parameters are set to $F=6$, $J=32$, $M=20$,
$\sigma_0=2$, and $\gamma=2$.
With these parameters, this model 
is in an unstable regime with the first Lyapunov exponent $\lambda_1 \simeq 0.84>0$.
The true initial condition $v_0 \in \mathbb{R}^M$ is given by model integration within an interval from a randomly chosen initial condition.
The observation is sampled from a model sequence starting from $v_0$,
with observational noise added to the sample.
The observation is defined only at the times $j+1=3,6,\cdots,30$ and space $l=1,2,\cdots,8$.
The first-guess $m_0$ is given by changing only the first component of $v_0$:
\begin{align}
(m_0)_1 = (v_0)_1+\sigma_0,~(m_0)_2 = (v_0)_2,\cdots,(m_0)_M = (v_0)_M.
\end{align}
The filter $R_k$ is in the form of Eq.\,(\ref{ex_Rk_even}) with
$j_k=10$ and $k=0.25^{-1}$.
$k$ is set so that the filtering term $\Delta S_k[\phi]$
is of order $1$, that is, 
$O(1)=\Delta S_k[\phi] \simeq k^2 \mathrm{var}(\phi)$,
where the typical fluctuation is assumed to be
$\mathrm{var}(\phi) \simeq k^{-2}
= 0.25^2$.
The time evolution of  (\ref{l96model})
and its adjoint are solved by the Runge--Kutta method
with time-step $\Delta t=0.1$.
The experiments are designed to
investigate how the action, the gradient of action, and
the effective gradient of action change
if we move the control variable as:
\begin{align}
\phi_1 = (v_0)_1+\sigma_0 \eta,~\phi_2 = (v_0)_2,\cdots,\phi_M = (v_0)_M,
\end{align}
where $-1 \le \eta \le 1$.

\begin{figure}
\begin{center}
\includegraphics[width=40em]{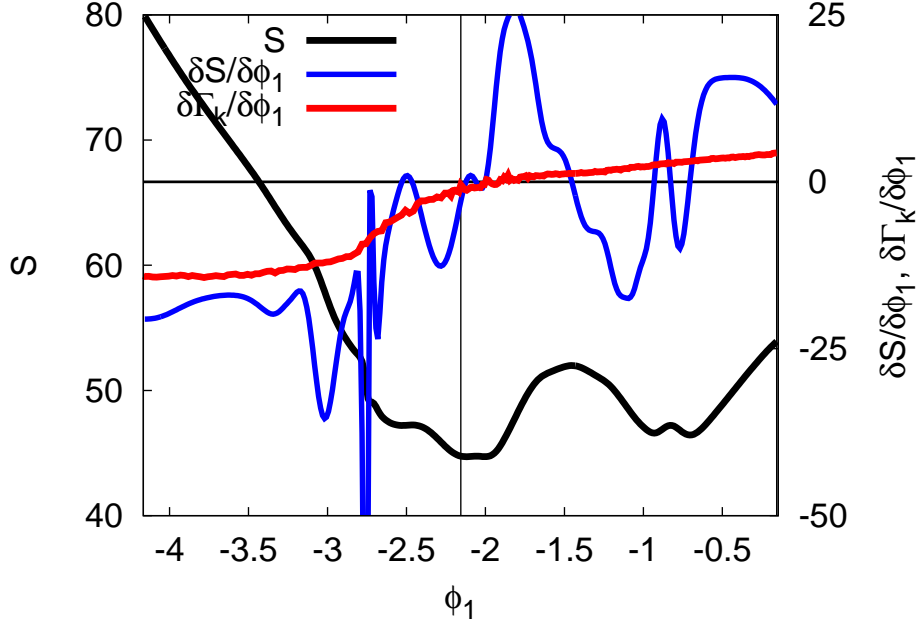}
\end{center}
\caption{
 Action $S[\phi]$ (black),
 gradient of action $\delta S[\phi]/\delta \phi_1$ (blue),
 and effective gradient of action $\delta \Gamma_k[\phi]/\delta \phi_1$ (red)
 for the Lorenz '96 model. The true value for the data
 assimilation problem is
 $\phi_1=-2.156$, and  the first-guess is $-0.156$.
}
\label{fig_l96_2}
\end{figure}

Fig.\,\ref{fig_l96_2} shows the
action (black), the gradient of action (blue), and the effective
 gradient of action (red) for this experiment.
It is clear that the original gradient has several zeros
that will complicate
the variational data assimilation using the
gradient.
However, the effective gradient has only one zero near the true value,
which is a similar result as for the Logistic map.

Thus, this example also shows the potential usefulness of the effective
gradient for finding the optimal initial condition.

\subsection{Slow and fast degrees of freedom}\label{slow_fast}
Toward the application to high-dimensional systems,
we briefly note a possible procedure for treating two distinct spatial modes
in the coarse-grained data assimilation.

Assume the infra-red filter has a sharp cutoff 
in the momentum representation:
\begin{align}
R_k(p) =
  \begin{cases}
  a^2 & \text{if} \; |p|<k \\
  0 & \text{otherwise} 
 \end{cases}, \label{Rk_slow_fast}
\end{align}
and the fluctuation $\chi$ can be decomposed into:
\begin{align}
\chi(p) =
  \begin{cases}
  \chi_s(p) & \text{if} \; |p|<k \\
  \chi_f(p) & \text{otherwise}
 \end{cases}.
\end{align}
Then, as the ``mass'' $a^2 \to \infty$, we have:
 \begin{align}
  \mathrm{e}^{-\Delta S_k[\chi_s+\chi_f]}
  =
  \mathrm{e}^{-\frac{a^2}{2}\chi_s^T\chi_s}
  &\to \delta[\chi_s],
 \end{align}
 which leads to:
\begin{widetext}
  \begin{align}
\frac{\delta \Gamma_k}{\delta \phi}[\phi]
&=
  \frac{
   \int \mathrm{d}\chi_s
   \int \mathrm{d}\chi_f
   \frac{\delta S}{\delta \phi}[\phi+\chi_s+\chi_f]
  \mathrm{e}^{
  -S[\phi+\chi_s+\chi_f]+S[\phi]+\frac{\delta \Gamma_k}{\delta
   \phi}[\phi](\chi_s+\chi_f)
  }
   \delta[\chi_s]}
  {
  \int \mathrm{d}\chi_s
  \int \mathrm{d}\chi_f
  \mathrm{e}^{
  -S[\phi+\chi_s+\chi_f]+S[\phi]+\frac{\delta \Gamma_k}{\delta \phi}[\phi](\chi_s+\chi_f)
  }
   \delta[\chi_s]}\\
   &\to
  \frac{
  \int \mathrm{d}\chi_f \frac{\delta S}{\delta \phi}[\phi+\chi_f]
  \mathrm{e}^{
  -S[\phi+\chi_f]+S[\phi]+\frac{\delta \Gamma_k}{\delta \phi}[\phi]\chi_f
  }
  }
  {
  \int \mathrm{d}\chi_f
  \mathrm{e}^{
  -S[\phi+\chi_f]+S[\phi]+\frac{\delta \Gamma_k}{\delta \phi}[\phi]\chi_f
  }
  }.
\label{defR3}
  \end{align}
\end{widetext}
   Hence, the sensitivity can be calculated
   as the average under the weight that
   integrates out the fast degrees of freedom,
   which will contribute to reducing the dimensionality of
   the path space.
   In the case of coupled atmosphere--ocean systems,
   we can assume that the atmospheric system is represented by $\chi_f$
   and the oceanic system is  $\chi_s$.
   This suggests that, through
   this coarse-graining procedure, the sensitivity regarding the coupled system can be
   expressed by the slow oceanic variables alone.

\subsection{Two-scale Lorenz '96 model}\label{L96_2}
 To illustrate the application to multiscale systems, as 
mentioned in Section \ref{slow_fast},
 we examine the sensitivity appearing
during data assimilation in the two-scale Lorenz '96 model \cite{l96}.
This system can be written 
 for the slow $(\zeta)$ and the fast $(\xi)$ variables 
as:
\begin{align}
 \frac{d\zeta_{l_1}}{dt} &= \zeta_{l_1-1} \left( \zeta_{l_1+1} - \zeta_{l_1-2} \right) -\zeta_{l_1}
 +F -\frac{h c}{b} \sum_{l_2=1}^{M_2} \xi_{l_2,l_1},\quad l_1=1,2,\cdots, M_1, \label{l96-2model-1}\\
 \frac{d\xi_{l_2,l_1}}{dt} &= c b \xi_{l_2+1,l_1} \left( \xi_{l_2-1,l_1} -
 \xi_{l_2+2,l_1} \right) -c \xi_{l_2,l_1} + \frac{h c}{b} \zeta_{l_1},\quad
 l_2=1,2,\cdots, {M_2}, \quad l_1=1,2,\cdots, M_1, \label{l96-2model-2}\\
\zeta_0 &= \zeta_{M_1}, \quad \zeta_{M_1+1} = \zeta_1, \quad \zeta_{-1} = \zeta_{M_1-1},\\
\xi_{0,l_1} &= \xi_{{M_2},l_1-1}, \quad 
\xi_{{M_2}+1,l_1} = \xi_{1,l_1+1}, \quad 
\xi_{{M_2}+2,l_1} = \xi_{2,l_1+1}, \quad l_1=1,2,\cdots, M_1,
\end{align}
where $\zeta(t) \in \mathbb{R}^{M_1}$,  $\xi(t) \in \mathbb{R}^{M_1 M_2}$.
We write the time evolution operator of the state
$\theta =(\zeta^T,\xi^T)^T$
from the initial condition
 as:
\begin{align}
\Psi^{(j)}\left[ \theta(t=0) \right]&=\theta(t=t_j),
\end{align}
with time-step $t_{j+1}-t_j=\Delta t$.
If we choose
the initial condition
$\theta(t=0) = \phi \in \mathbb{R}^M$,
where $M=M_1 + M_1 M_2$, as
 the control variable, we can define a cost function as follows:
\begin{align}
 S[\phi] &= 
\frac{1}{2\sigma_0^2} \|\phi -m_0 \|^2 +
\sum_{j=0}^{J-1} \frac{1}{2\gamma^2}\left\|y_{j+1} -
 \Psi^{(j+1)}[\phi]\right\|^2, \label{cost_twoscale_l96}
\end{align}
where $\|\cdot\|$ is the Euclidean norm.
The parameters are set to
$F=6$, $J=110$, $M_1=5$, $M_2=3$,  $h=1.6$, $b=c=10$, 
$\sigma_0=2$, and $\gamma=2$.
With these parameters, this model 
is in an unstable regime with the first Lyapunov exponent $\lambda_1
\simeq 9.0>0$.
The true initial condition $v_0 \in \mathbb{R}^M$ is given by model integration within an interval from a randomly chosen initial condition.
The observation is sampled from a model sequence  starting from $v_0$,
with observational noise added to the sample.
 The observation is defined only at the times $j+1=3,6,\cdots,108$ 
 and for the slow variables
$l_1=1,2,\cdots,M_1$.
The first-guess $m_0$ is given by changing only the first component of $v_0$:
\begin{align}
(m_0)_1 = (v_0)_1+\sigma_0,~(m_0)_2 = (v_0)_2,\cdots,(m_0)_M = (v_0)_M.
\end{align}
Similar to Eq.\,(\ref{Rk_slow_fast}), we set 
the filter $R_k$ as
\begin{align}
R_k &=
  \begin{cases}
  k_{\zeta}^2  & \text{for} ~~\zeta(t=0) \\
  k_{\xi}^2    & \text{for} ~~\xi(t=0)
 \end{cases}, \label{Rk_twoscale}
\end{align}
with $k_{\zeta}=0.01^{-1}$ and $k_{\xi}=0.04^{-1}$. 
This setting,  $k_{\zeta} > k_{\xi}$,
  mainly integrates out the 
fast degrees of freedom $\xi(t=0)$,
whose typical fluctuation is assumed
to be
$\mathrm{var}\left(\xi(t=0)\right) \simeq k_{\xi}^{-2} 
= 0.04^2$.
The time evolution of (\ref{l96-2model-1}) and (\ref{l96-2model-2})
as well as their adjoints are solved by the Runge--Kutta method
with time-step $\Delta t=0.006$.
As in section \ref{L96_1}, the experiments are designed to
 investigate the changes in the action, the gradient of action, and
 the effective gradient of action if we move the control variable as:
\begin{align}
\phi_1 = (v_0)_1+\sigma_0 \eta,~\phi_2 = (v_0)_2,\cdots,\phi_M = (v_0)_M,
\end{align}
where $-1 \le \eta \le 1$.

\begin{figure}
\begin{center}
\includegraphics[width=40em]{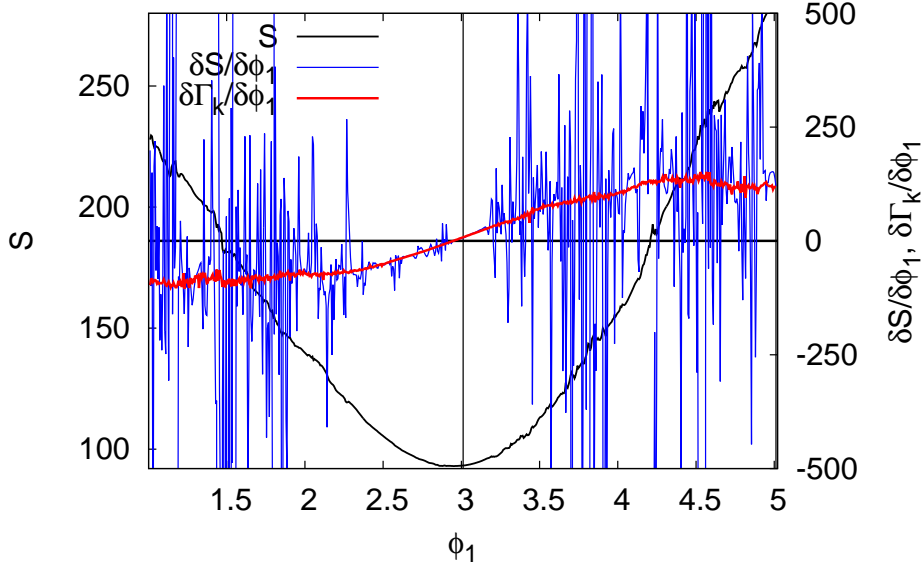}
\end{center}
\caption{
 Action $S[\phi]$ (black),
 gradient of action $\delta S[\phi]/\delta \phi_1$ (blue),
 and effective gradient of action $\delta \Gamma_k[\phi]/\delta \phi_1$ (red)
 for the two-scale Lorenz '96 model.
 The true value for the data
 assimilation problem is
 $\phi_1=3.011$, and  the first-guess is $5.011$.
\label{fig_twoscale_l96}
}
\end{figure}

Fig.\,\ref{fig_twoscale_l96} shows the
 action (black), the gradient of action (blue), and the effective
 gradient of action (red) for this experiment.
The action 
 is almost parabolic in shape with many small bumps
 due to the fast degrees of freedom.
 Consequently, the gradient has many zeros 
 that can complicate
the variational data assimilation using it.
However, the effective gradient has only one zero near the true value,
 similar to the result obtained for the Logistic map and for the one-scale
Lorenz '96 model.
 Thus, this example shows the potential application of the effective
gradient for finding the optimal initial condition in multiscale data
assimilation.

 \section{Conclusion}
 We have investigated the use of the
 effective average action and its gradient in
 multiscale data assimilation.
A framework
has been proposed
 that allows the multiscale data assimilation problem to be solved
by replacing the cost function with the effective average action.
We have also proposed
a novel method of evaluating the gradient of
the effective average action numerically.
In principle, this enables  a broader range of data
assimilation problems to be solved
by seeking the stationary point of this effective average action
 numerically using its gradient.

 This work can be summarized as follows.
\begin{itemize}
\item The concept of
the effective average action provides a consistent framework for data
      assimilation in nonlinear multiscale systems.
\item
If we can numerically evaluate the path integral:
\begin{align*}
 \frac{\delta \Gamma_k}{\delta \phi}[\phi]
 &=
\frac{\int \mathrm{d}\chi \frac{\delta S}{\delta \phi}[\phi+\chi] \mathrm{e}^{-\mathfrak{R}[\phi,\chi]}}
 {\int \mathrm{d}\chi \mathrm{e}^{-\mathfrak{R}[\phi,\chi]}}
  \equiv
 \left< \frac{\delta S}{\delta \phi}[\phi+\chi] \right>_{\mathfrak{R}},\\
\mathfrak{R}[\phi,\chi]&\equiv
S[\phi+\chi]-S[\phi]-\frac{\delta \Gamma_k}{\delta \phi}[\phi] \chi
+\Delta S_k[\chi]
\end{align*}
 with reasonable accuracy and computational burden,
then we obtain the coarse-grained sensitivity, which constitutes
 a key factor in the variational data assimilation of a coarse-grained field.

\item
     The proposed  procedure
  estimates the coarse-grained sensitivity
through the Metropolis method by averaging an ensemble of
     original sensitivities that are distributed
     according to weights related to the nonlinearity:

\begin{align*}
 \left< \frac{\delta S}{\delta \phi}[\phi+\chi] \right>_{\mathfrak{R}}
&\simeq
\frac1N \sum_{n=1}^{N} \frac{\delta S}{\delta \phi}[\phi+\chi^{(n)}].
\end{align*}

 \item The stationary problem for the effective average action $\Gamma_k[\phi]$
       can be solved using a gradient method with the gradient
       $\delta \Gamma_k[\phi]/\delta \phi$.

\item The stationary value $\hat{\phi}$
      represents the extremum for the coarse-grained field
      after integrating out the ultra-violet
      fluctuations.
      This can be regarded as a solution of the multiscale data assimilation
      problem.
\item We demonstrated the usefulness of the effective gradient for data
      assimilation in a simple setting with the double-well potential,
       the Logistic map, the one-scale Lorenz '96 model,
      and the two-scale Lorenz '96 model.
\end{itemize}
Future research should consider the following issues.
\begin{itemize}
 \item The infra-red filter works well
	 when we can separate slow and fast modes cleanly, as described in
	 sections \ref{slow_fast} and \ref{L96_2}. However, in general, we have to
	 deal with control variables that have continuous spectra.
	 For such cases, we should carefully consider 
	 an infra-red filter design that is suitable for
        revealing the slow dynamics of the system under consideration.
 \item In the case of a larger system,
       the computational burden of the Metropolis method could be huge,
       because we require many samples to yield a statistically
       reasonable integration result. Moreover,  this should be
       incorporated into a recursive procedure,
       or a fixed point calculation.
\end{itemize}
Despite these technical difficulties,
the coarse-grained sensitivities
are of great importance,
since they provide an invaluable
 perspective on the slow dynamics
 of multiscale systems.
     It should be noted that our approach in the present form has
     a fairly limited scope of application to data assimilation problems
      in geoscience, which typically require more high-dimensional systems.

\acknowledgments
The author is grateful to H. D. I. Abarbanel
for motivating this research and
for several suggestions.
The numerical simulations were
performed on JAMSTEC SC supercomputer system.
\appendix
\section{Properties of $\Gamma_k$}
\subsection{Perturbation expansion}\label{app1}
To evaluate the path integral (\ref{defGamma3})
 through a perturbation expansion,
we apply the approximation
$\delta \Gamma_k[\phi]/\delta \phi \simeq \delta S[\phi]/\delta
\phi$ in the exponent of Eq.\,(\ref{defR}),
and truncate the Taylor series expansion of $S[\phi+\chi]$
to the quadratic order:
\begin{align}
&S[\phi+\chi]-S[\phi]-\frac{\delta \Gamma_k}{\delta \phi}[\phi] \chi
+ \Delta S_k[\chi]\nonumber\\
&\simeq
S[\phi]
+\frac{\delta S}{\delta \phi}[\phi] \chi
+\frac12 \chi^T \frac{\delta^2 S}{\delta \phi^2}[\phi] \chi
-S[\phi]-\frac{\delta S}{\delta \phi}[\phi] \chi
+ \frac12 \chi^T R_k \chi \nonumber \\
&=
\frac12 \chi^T \frac{\delta^2 S}{\delta \phi^2}[\phi] \chi
+ \frac12 \chi^T R_k \chi
=\frac12 \chi^T \left(
\frac{\delta^2 S}{\delta \phi^2}[\phi]+R_k
\right) \chi.
\end{align}
Taking the Gaussian integral, we obtain
\begin{align}
\Gamma_k[\phi]
&\simeq S[\phi]
-\log{\int \mathrm{d}\chi
\mathrm{e}^{-\frac12 \chi^T \left(\frac{\delta^2 S}{\delta
 \phi^2}[\phi]+R_k\right) \chi}}\nonumber\\
&=S[\phi]+\frac12 \log{\det{\left(\frac{\delta^2 S}{\delta
 \phi^2}[\phi]+R_k\right)}}.\label{defGamma2}
\end{align}
Thus, we require at least the second derivative of the action $S[\phi]$ for
the perturbation calculation of $\Gamma_k[\phi]$.

The same procedure can be applied to the gradient (\ref{eaa_en}) as follows:
\begin{align}
 \frac{\delta \Gamma_k}{\delta \phi}[\phi]
&\simeq
\frac{\delta S}{\delta \phi}[\phi]
+
\frac12 \left<
\chi^T \left(\frac{\delta^3 S}{\delta \phi^3}[\phi]\right) \chi
\right>_{\mathfrak{R}}\nonumber\\
&=
\frac{\delta S}{\delta \phi}[\phi]
+
\frac12 \mathrm{tr}\left\{
\frac{\delta^3 S}{\delta \phi^3}[\phi]
\left(
\frac{\delta^2 S}{\delta \phi^2}[\phi]+R_k
\right)^{-1}
\right\}.\label{gradient2}
\end{align}
Hence, we require at least the second and third derivatives of the action $S[\phi]$ for
the perturbation calculation of the gradient of $\Gamma_k[\phi]$.

\subsection{Relationship to the growth of instabilities}\label{app2}
We consider the case where $S[\phi]$ is the cost function of
strong-constraint 4D-Var  \cite{QJ:QJ36}.
The meaning of $\log{\det{}}$ in Eq.\,(\ref{defGamma2}) can be clarified
by considering the basis of singular vectors.
With the singular values $\sigma_1>\sigma_2>\cdots$, we can write:
\begin{align}
 \frac{\delta^2 S}{\delta \phi^2}[\phi]
&\simeq
\left( \frac{\delta F}{\delta \phi}[\phi] \right)^T
\left( \frac{\delta F}{\delta \phi}[\phi] \right)
=\mathrm{diag}{[\sigma_1^2,\sigma_2^2,\cdots]}.
\end{align}
Using a positive constant $k \gg 1$, we can define the infra-red filter as:
\begin{align}
R_k =
  \begin{cases}
  k^2 & \text{if} \; \sigma_i<k \\
  0 & \text{otherwise}
 \end{cases}.
\end{align}
Then, we have
\begin{align}
 \frac{\delta^2 S}{\delta \phi^2}[\phi]+R_k
&\simeq
\mathrm{diag}{[\sigma_1^2,\sigma_2^2,\cdots,k^2,\cdots,k^2]},\\
\frac12 \log{\det{\left(
 \frac{\delta^2 S}{\delta \phi^2}[\phi]+R_k
\right)}}
&\simeq
\sum_{\sigma_i \ge k}\log{\sigma_i}
+
\sum_{\sigma_i < k}\log{k}.\label{sigma}
\end{align}
That is, the term $\log{\det{}}$ represents the sum of the logarithms of
the leading singular values.
The additional term in Eq.\,(\ref{sigma}) has the effect
of integrating out the growing disturbances in the cost function.

\section{An algorithm for coarse-grained data assimilation}\label{app3}
   \begin{algorithm}[H]
    \caption{coarse-grained\_data\_assimilation}
   \begin{algorithmic}[0]
    \State {$\Psi_0 \leftarrow v_0$}\Comment{set true value}
    \For {$t=0 \to T-1$}
   \State {$\Psi_{t+1} \leftarrow \psi(\Psi_{t})$}
   \State{generate $\xi \sim \mathcal{N}(0,I)$}
   \State {$y_{t+1} \leftarrow \Psi_{t+1}+\gamma \xi$}  \Comment{set observation}
    \EndFor
    \State {$\phi_0 \leftarrow m_0$} \Comment{set first-guess}
    \For{$i=0 \to I-1$} \Comment{assimilation loop}
    \State \Call{calc\_cost}{$y_{1:T},\phi_i,\Psi_{0:T},S_i$}
    \State \Call{calc\_sensitivity}{$y_{1:T},\Psi_{0:T},\nabla S_i$} \Comment{See gradient\_calculations}
    \State \Call{calc\_coarse-grained\_sensitivity}{$y_{1:T},\phi_i,S_i,\nabla S_i,\nabla \Gamma_i$} \Comment{See gradient\_calculations}
    \If{$| \nabla \Gamma_{i} |< C_{\text{th}}$}
    \State \Return{$\phi_i,S_i$}
    \EndIf
    \State {$\phi_{i+1} \leftarrow \phi_{i}-\alpha \nabla
    \Gamma_{i}$}\Comment{update control variable}
    \EndFor
    \Statex
   \Procedure{calc\_cost}{$y_{1:T},\phi_i,\Psi_{0:T},S_i$}
   \State {$\Psi_0 \leftarrow \phi_i$} \Comment{background term}
    \State{$S_i \leftarrow \frac1{2 \sigma_0^2}| \Psi_0-m_0 |^2$}
   \For {$t=0 \to T-1$}
   \State {$\Psi_{t+1} \leftarrow \psi(\Psi_{t})$}\Comment{forward time stepping}
   \State {$S_i \leftarrow S_i + \frac1{2 \gamma^2}
    |y_{t+1}-\Psi_{t+1}|^2$}  \Comment{observational term}
    \EndFor
    \EndProcedure
    \end{algorithmic}
    \end{algorithm}
    \begin{algorithm}[H]
    \caption{gradient\_calculations}
    \begin{algorithmic}[0]
   \Procedure{calc\_sensitivity}{$y_{1:T},\Psi_{0:T},\nabla S_i$}
    \State {$\widehat{\Psi}_{0:T} \leftarrow 0$}
   \For {$t=T-1 \to 0$}
    \State {$\widehat{\Psi}_{t+1} \leftarrow
    \widehat{\Psi}_{t+1}-\frac1{\gamma^2} (y_{t+1}-\Psi_{t+1})$} \Comment{observational term}
    \State {$\widehat{\Psi}_t \leftarrow \widehat{\Psi}_t
    + \left(\frac{\partial \psi}{\partial \Psi_t}\right)^T
     \widehat{\Psi}_{t+1}$} \Comment{adjoint time stepping}
    \State {$\widehat{\Psi}_{t+1} \leftarrow 0$}
    \EndFor
    \State{$\widehat{\Psi}_0 \leftarrow \widehat{\Psi}_0
     +\frac1{\sigma_0^2}(\Psi_0-m_0)$} \Comment{background term}
   \State {$\nabla S_i \leftarrow \widehat{\Psi}_0$}
      \State {$\widehat{\Psi}_0 \leftarrow 0$}
    \EndProcedure
    \Statex
    \Procedure{calc\_coarse-grained\_sensitivity}{$y_{1:T},\phi_i,S_i,\nabla S_i,\nabla \Gamma_i$}
    \State{$\nabla \Gamma_i \leftarrow \nabla S_i$}\Comment{first-guess
     of gradient in weight}
   \For {$l=1 \to L$} \Comment{successive correction of gradient in weight}
    \State{$\chi \leftarrow 0$}
    \State{$\nabla S \leftarrow \nabla S_i$}
    \State{$R \leftarrow 0$}
    \State{$\nabla R \leftarrow 0$}
    \State{$\nabla \Gamma^{\text{acc}} \leftarrow 0$}
   \For {$n=0 \to N-1$} \Comment{Markov-chain loop}
    \State{generate $\xi \sim \mathcal{N}(0,I)$}
    \State{$\chi^* \leftarrow \chi-\frac12 \sigma^2 \nabla R+\sigma
     \xi$}\Comment{proposed}
    \State \Call{calc\_cost}{$y_{1:T},\phi_i+\chi^*,\Psi_{0:T},S^*$}
    \State \Call{calc\_sensitivity}{$y_{1:T},\Psi_{0:T},\nabla
     S^*$}
    \State{$R^* \leftarrow S^*-S_i
    -\left<\nabla \Gamma_i, \chi^* \right>+\frac12 k^2 | \chi^* |^2$}
    \State{$\nabla R^* \leftarrow \nabla S^*
    -\nabla \Gamma_i+k^2 \chi^*$}
    \State{$q_+ \leftarrow \frac{1}{2\sigma^2} |\chi-\frac12 \sigma^2
     \nabla R-\chi^* |^2$}
    \State{$q_- \leftarrow \frac{1}{2\sigma^2} |\chi^*-\frac12  \sigma^2 \nabla R^*-\chi |^2$}
     \State{$a \leftarrow -R^*+R-q_-+q_+$}
    \State{generate $\zeta \sim \mathcal{U}(0,1)$}
    \If{$a \geq 0$ or $\zeta < \exp{(a)}$} \Comment{Metropolis criterion}
    \State {$\chi \leftarrow \chi^*$}
    \State{$\nabla S \leftarrow \nabla S^*$}
    \State{$R \leftarrow R^*$}
    \State{$\nabla R \leftarrow \nabla R^*$}
    \EndIf
    \State{$\nabla \Gamma^{\text{acc}} \leftarrow  \nabla
     \Gamma^{\text{acc}} +\nabla S$} \Comment{accumulate gradient}
   \EndFor
    \State{$\nabla \Gamma_i \leftarrow \nabla \Gamma^{\text{acc}} /
     N$}\Comment{take average}
     \EndFor
     \EndProcedure
    \end{algorithmic}
   \end{algorithm}

\section*{References}

\bibliographystyle{apsrev4-1} 
\bibliography{./ref} 
\end{document}